\documentclass[12pt,a4paper,aps]{article}
\usepackage{a4}
\usepackage{epsfig}
\usepackage[hmargin=2cm,vmargin=2cm,nohead]{geometry}

\usepackage{graphicx}
\usepackage{amssymb}
\usepackage{epstopdf}
\newif\ifpdf
  \ifx\pdfoutput\undefined
  \pdffalse
\else
  \pdfoutput=1
  \pdftrue
\fi

\ifpdf
  \DeclareGraphicsExtensions{.pdf,.jpg,.JPG,.jpeg,.gif}
\else
  \DeclareGraphicsExtensions{.ps,.eps,.eps.gz,.ps.gz}
\fi

\ifpdf
\usepackage[colorlinks=true, pdfstartview=FitV, linkcolor=blue, 
            citecolor=blue, urlcolor=blue]{hyperref}
\hypersetup{pdfstartview={Fit -32768},
  pdfpagemode=UseOutlines,
  plainpages=true,
  bookmarksnumbered=true,
  bookmarksopen=false
  }

\begin{document}

%\begin{center}
%{\huge \bf 
\title{Regional Oil Extraction and Consumption:  \\
A simple production model for the next 35 years\\
Part I} 

%\end{center}

\author{
Michael Dittmar\thanks{e-mail:Michael.Dittmar@cern.ch},\\
Institute of Particle Physics,\\ 
ETH, 8093 Zurich, Switzerland\\
\date{\today} 
}
\maketitle
%\par
%\begin{center}
%{\large \bf 
%\author{Michael Dittmar \\ 2015}
%\end{center}

\begin{abstract}
\noindent 
%Essentially nobody disagrees with the ideas that (1) oil is a finite energy resource and that 
%(2) despite huge scientific, technological and political efforts, reduce the oil dependence of the richer %countries 
%and of the global economy, no realistic alternatives especially for the oil based transport system have been %found so far. 
The growing conflicts in and about oil exporting regions and speculations about volatile oil prices during the last decade have renewed the public interest in predictions for the near future oil production and consumption. 
Unfortunately, studies from only 10 years ago, which tried to forecast the oil production during the next 20-30 years, failed to make accurate predictions for today's global oil production and consumption.
Forecasts using economic growth scenarios, overestimated the actual oil production,  
while models which tried to estimate the maximum future oil production/year,
using the official country oil reserve data, predicted a too low production.   

\noindent
In this paper, a new approach to model the maximal future regional and 
thus global oil production (part I) and consumption (part II) during the next decades is proposed.  

\noindent
Our analysis of the regional oil production data during past decades shows that, in contrast to periods when production was growing and growth rates varied greatly from one country to another, remarkable similarities are found during the plateau and decline periods of different countries.
As a result we propose to model the oil production decline phase in all oil producing regions
with a universal decline function which assumes that a few years with a stable production, 
will be followed by about 5 years with a 3\%/year decline and 6\%/year afterwards. 
This plateau and decline function was found to describe 
accurately the oil production of the past 15 years from Western Europe (EU+Norway), Mexico and Indonesia. 

\noindent
Following this model, the particular production phase of each major oil producing country and region is determined essentially only from the recent past oil production data. Accordingly, we find that the conventional oil production in all regions, outside of the Middle East OPEC countries, will start the 3\%/year decline phase between the years 2015 and 2020. Using these data, the model is then used to predict the production from all major oil producing countries, regions and continents up to the year 2050. The limited regional and global potential to  
compensate this decline with unconventional oil and oil-equivalents is also presented.

\end{abstract}
%%%%%%%%%%%%%%%%%%%%%%%%%%%%%%%%%%%%%%

\newpage
\section{Introduction: modelling the future oil production}

It is difficult to imagine a modern industrialised way of life and the globalised economy without oil. 
For distances from a few tenth to thousands of km, oil is the ideal energy career for the transport of people and all kinds of consumer goods, including nuclear power plants, windmills and photovoltaics.
Oil is also essential for the functioning of today's Cities and Megacities, which can not be fed and maintained 
without an oil based globalised industrialised agriculture and the subsequent transport of the produced food.

Despite huge scientific and technological efforts, combined with the strong wish to reduce the oil dependence of the rich and highly industrialised countries during the last 30-40 years, no realistic alternatives have been found so far. To the contrary, the growing interdependence of the global economy, based on the cheap oil based transport system, only increased this oil dependence.

As it is generally accepted that oil is a finite resource and that the future global production and consumption must eventually decline, the future crude oil production possibilities are thus important research topics.  One finds that essentially two different types of modelling for the future oil production and consumption exist.  

The first type of forecasts are the ones from government agencies or transnational oil companies, which 
are inspired by political and economical regional and global ideas about the expected 
future demand growth.  Examples are the annual energy outlooks from the IEA \cite{IEAWEO}, the International Energy Agency of the OECD countries, and from the EIA \cite{EIAfuture}, the Energy Information Agency of the USA government, or the regular energy outlooks from oil companies like BP \cite{BPfuture}. 
One finds that those global studies from the past decade(s) assumed that the remaining conventional and unconventional oil reserves would be cheap enough to allow economic growth trends observed during previous decades to continue. For example, the 2001 (WEO 2001) version of the annual energy outlook from the IEA \cite{IEA2001}, imagined that global oil production and consumption would continue to grow at least to the year 2020 by about 2\%/year. Accordingly, those economic inspired models assumed, that reserves are sufficient to guarantee that the global consumption of oil and oil-equivalent liquids will grow from about 80 mbd (million barrel per day) around the year 2000 to about 105 mbd in the year 2015 and further to 115 mbd by 2020. Such production and consumption growth was assumed to be possible with a relatively stable oil price of less than 30 dollar/barrel. 
The reality was very different as the oil price increased steadily to about 100 dollar/barrel in 2007/8.
During the 2008 global economic crisis and during the following years the oil price fluctuated between 
50 to 100 dollars/barrel, about a factor of 2 to 4 higher than predicted.
Using the August 2015 IEA estimate \cite{IEAoil2015}, the 2015 oil-equivalent production and consumption is estimated to be about 93-94 mbd, corresponding to an increase of about 13 mbd with respect to the year 2000. This is a factor of two lower than the one from the WEO 2001 report \cite{IEA2001},  
where an increase of 25 mbd by 2015 was predicted. 

Instead of trying to change the underlying wrong assumptions,   
subsequent economically based forecasts simply reduced the annual consumption growth rate and extended the forecast period.  For example the global production and consumption of all oil-equivalent liquids for the year 2020 decreased from 115 mbd (WEO 2001) to about 109 mbd (WEO 2006) \cite{IEA2006} and 96 mbd (WEO 2015) in the latest report \cite{IEA2015}. 
It is obvious that the methods used by these economically inspired models did not lead to reliable forecasts beyond a few years. 

The other type of forecasting models is based on educated guesswork about the remaining exploitable oil reserves, combined with assumptions about future extraction technologies. 
Those geophysical inspired approaches result essentially always in estimates about the maximum possible global annual oil peak production followed by the subsequent declining production.
For example, the oil geologists C. Campbell and J. Laherr\`ere estimated in their 1998 Scientific~American
article, ``The End of Cheap Oil" \cite{scientificAM}, that the global crude oil production would reach a maximum with a production of about 71 mbd (26 Giga barrel of oil per year) around the year 2005.
After 2005 the production would decline by 2-3\% per year, corresponding to a production of 
about 60 mbd by 2015. This 1998 publication was based on the uncertain official regional oil reserves, corrected by their reevaluation of some of these numbers and some assumptions about possible future discoveries. 
Many more such reserve based regional and global production forecasts \cite{ASPO}, \cite{oildrum}, were published during the next few years and essentially all of them predicted a global conventional oil production peak between 2005 and 2010, followed by a decline of a few percent a year.

However, even if today's important contributions from unconventional oil,  
and other oil-equivalent liquids are excluded, the global conventional oil production\footnote{In the following we use the crude and condensate oil production numbers as given by the EIA database \cite{EIAoildata} minus the
contribution coming from tight oil and tar sands in North America.}
neither 
grew nor declined since 2005 but has remained essentially at a stable plateau of 
about 72 mbd \cite{oilage1}. 

As can be seen from Table 1, large differences between the predicted and real production 
trends in both types of forecasting methods are observed for Russia, China and the Middle East OPEC countries. 

{\tiny
\begin{table}[h]
\begin{center}
\begin{tabular}{|c|c|c|c|c|c|}
\hline
Forcast/Real   & region  & prod./mod. & prod./mod.  & prod./mod.&modelled \\
oil production   &             & [mbd]       & [mbd]       & [mbd]                    &   [mbd]  \\
            &             & 2005        &  2010       &  2015(data 2014) & 2020   \\ 
\hline
all liquids real     &  World  all               &  85           & 88                &   93       &     (?)   \\
IEA 2001/06       &  World all                &  87 / 84    &  96 / 91        &   105 / 99 & 115 / 106               \\
ASPO 2001/06   &  World all                &  78 / 80    &  83 / 90        &    75 / 85 &   67 / 75               \\
\hline
conv. oil real      &  World conv.        &  72.2       & 72.4           &   71       &     (?)   \\
IEA 2001/06      &  World conv.        &  74 / 71    &  80 / 76        &     86 / 80 &   93 / 84.5               \\
ASPO 2001/06  &  World conv.        &  66 / 70.5 &  69 / 74        &     61.5 / 66 &   51.5 / 56               \\
\hline
conv. oil real     & OECD Europe    & 5.1         &   3.7           &   2.8      &    (?)    \\
IEA 2001/06     & OECD Europe    &   6.0 / 4.8            &   5.2 / 3.8         &  4.5 / 2.9    &   3.5 / 2.3     \\
ASPO 2001/06 & OECD Europe    &  5 / 5    &  3.6 / 3.4        &     2.8 / 2.3 &   1.9 / 1.6               \\
\hline 
 conv. oil real                   & Russia                &  9.0        &   10.1         &  10.1     &     (?)   \\
 IEA 2001/06                   &Russia                &     6.7 / 9.2           &   7.1 / 10.5  &  7.5 / 10.6   & 7.9 / 11  \\   
ASPO 2001/06  & Russia    &   8.4 / 9.2    &  6.7 / 8.5        &     5.5 / 6.9 &   4.3 / 5.7               \\
\hline 
conv. oil real          & China  &  3.6        &  4.1             &  4.2       &     (?)     \\
IEA 2001/06          &China   &  3.2 / 3.6  &  3 / 3.8          &  2.8 / 3.7 &  2.6 / 3.2       \\
ASPO 2001/06 & China       &  3.0 / 3.4    &  2.4 / 2.6         &  2.0 /2.15 &   1.6 / 1.7                \\
\hline
conv. oil real   & OPEC ME &  22.4          & 24.5             &  25       &       (?)   \\  
IEA 2001/06   & OPEC ME  &  25 / 21      &  30.5 / 22       &   37 / 25.7      & 46.7 / 34.5       \\
ASPO 2001/06 & OPEC ME  &  17 / 20   &  24.4 / 19         &     23.3 / 19 &   22.3 / 19               \\      
\hline
\end{tabular} %\vspace{0.1cm}
\caption{Production of all liquids (world) and production of conventional oil (world and different regions),\cite{EIAoildata},  and the modelled production from the IEA (WEO 2001 and WEO 2006)
and from the ASPO Newsletter Ireland (February 2002 and November 2006) \cite{ASPO}. OPEC ME stands for the Middle East OPEC countries. 
}
\end{center}
\end{table}
}
For example, the production of all oil-equivalent liquids in 2014 reached about 10.5 mbd in Russia 
and 4.3 mbd in China \cite{EIAoildata}. In comparison, the ASPO 2006 study predicted production peaks around the year 2010 of about 9.2 mbd and 3.4 mbd for Russia and China respectively and a decline 
to about 7 mbd (Russia) and 2 mbd (China) around the year 2015.
It is interesting to note that the economically inspired IEA (WEO 2001) global growth scenario 
predicted similar production peaks and declines for Russia and China. 
In contrast, very different production forecasts were made for the Middle East OPEC countries.
The IEA essentially always expected that the production in this region would increase by  
about 10\% every 5 years, while the resource based models predicted essentially  
a flat production between 2010 and 2020. In reality, the production has up to now increased, but only half as much 
as predicted by the IEA.  
  
Despite these failures, it is important to acknowledge that the 1998 paper from C. Campbell and 
J. Laherr\`ere \cite{scientificAM}, and the subsequent resource based studies predicted essentially correctly, the end of cheap oil  (below 30 dollars/barrel) and the steeply declining oil production in Western Europe. 
Perhaps one might also accept the explanation that the failures of the resource based forecasts 
originate from the inaccurate public oil reserve data and not from the 
method used. Unfortunately, one can not expect to improve the accuracy of those forecasts, 
as the officially given reserves, regularly published from the EIA \cite{EIAoildata} or private companies like from BP, \cite{BPfuture}, are a badly defined mixture between the remaining, or original in place, exploitable conventional oil with different qualities, unconventional tight oil and oil from tar sands. Furthermore these reserves are also biased, either upwards or downwards, by the political and economic interests of oil companies and entire countries. 

In the following, we first (section 2) introduce our basic model assumptions using the past conventional oil production data from Western Europe (EU+Norway), Indonesia and Mexico. 
Using this model, the maximal possible regional conventional oil production is estimated for all oil producing region and continents and up to the year 2050 (section 3.1-3.7). {\it It is important to consider that the 
real production might be considerable lower, as production interruptions because of wars or 
because of regional or global economic recessions during the coming years and decades are not
included in this approach.} Possible reasons for a steeper production decline are discussed for example in more detail in reference \cite{Gail}. 

The prospects of substituting unconventional oil, and supposedly oil-equivalent liquids, for crude oil are discussed in section 3.8. In section 3.9 the model is tested using 
the observed oil production trend between the years 2000-2005 to make 
predictions for 2014 which can be compared with the actual 2014 crude oil production.
In section 3.10 we compare the results from our 
global oil production forecast for the next decades with the ones from other estimates.

In section 4, the consequences of the modelled near future oil and oil-equivalent production for the different consumer regions around the year 2020 are discussed. A more detailed analysis of the future import and export possibilities of the different larger producer and consumer countries and regions will be presented in a subsequent paper. 

\section{A simple model to forecast the oil production over the next decades}

As presented in the introduction, oil reserve data are not presented in  
a transparent scientific format which would include uncertainties.
Unfortunately, the public reserve data do not exist in such a format, neither for individual oil fields nor 
for entire countries or regions. As a consequence, models which are based on resource data alone, 
can only lead to accurate quantitative predictions if the used reserve data are accurate.

This is also true for perhaps the most transparent official reserve data presented in the 
past from the UK and Norway. For example, the crude oil reserves of the UK and Norway in 1983 were given as 14 Gb (Giga barrel of oil (= $10^{9}$ barrel)) and 6.8 Gb respectively. 

The original in place reserves can also be estimated by combining: (a) the total oil extraction between 1984 to 2013, which corresponds,  according to the EIA data \cite{EIAoildata}, to 21.4 Gb (UK) and 24 Gb (Norway) and  
(b) the remaining reserves at the end of 2013 are 3.1 Gb (UK) and 5.4 Gb (Norway).  The resulting 1983 
reserves were about 25 Gb (UK) and 30 Gb (Norway), and thus much larger than originally claimed. 

For Mexico \cite{EIAoildata}, one finds that a total of 23 Gb had been extracted between 1994 and 2013 and 10 Gb remain for extraction. Consequently the reserves, given as 51 Gb in 1993, were overestimated by 
18 Gb. It follows that estimates for the future oil extraction from these three countries, based only 
on the officially reserves given in 1983 and 1993, would not have resulted in a prediction anywhere near 
the real and declining production during the last 5-10 years. 

Knowing that the reserve data for many other countries are even less transparent than in the examples above, it appears impossible to make accurate oil production forecasts based on official reserve data. 
Perhaps these problems explain already why the past predictions failed to describe the actual oil production for Russia, China and the Middle East. 
Furthermore, as improvements of exploration technologies and the 
associated monetary exploration costs are somewhat a secret of the oil companies, 
it is also clear that the future production from unconventional oil and oil-equivalent liquids are also 
very uncertain.   
 
Despite these uncertainties, it is undoubted that oil is a finite fossil fuel, and that oil production, depending on the remaining regional reserves, must eventually decline in all regions of the planet. The question is thus if 
an oil production model with an improved predictive power can be found. 

In the following we propose a new and simple model, which is essentially based 
only on the past production data from different countries.
These data are easily available for everyone and are at the same time believed to be accurate  \cite{EIAoildata}. 

Additionally, it is assumed that the regional oil production depends not only on the amount of exploitable oil reserves in the territory, but also on the possibility of transporting the oil from the producing region to the user region and this under the most profitable conditions for the private or public owner of the field and for the exploiting companies. Furthermore, we assume that the highest profits can be made somewhere between these conditions: 

\begin{itemize} 
\item The oil production in a particular region rises as quickly as technically possible to a maximum, which can be maintained only for a few years, because the overall production must decline once new fields can not be opened as quickly as the older fields decline. 
The exploitation of the oil fields in the North Sea, mainly operated from the UK and Norway, is a good example for such regions. 
\item The oil reserves in the country are huge, and sufficiently large enough profits can be made from exports, even if the oil production is limited for political or economic reasons below the possible maximum. 
The oil production in Saudi Arabia and its oil rich neighbouring countries can be considered as examples.
\item Production can also be limited, if the theoretically producible oil in a given region can not be exported easily because either there are no pipelines nor does a sufficiently large oil tanker based transport capacity exist.  The oil production in a landlocked country like Kazakhstan is an example for this case. 
\end{itemize}

Furthermore, we assume that the consequence of finite exploitable oil reserves, 
within a given field and region, will eventually result in a steeply and on average universally declining oil production. Support for such a universal decline comes from the IEA studies in the World Energy outlooks 2008 and 2013, which concluded that older oil fields eventually all decline on average by about 6\%/year \cite{IEAWEO}. 
The authors of the IEA study (2013) have found this number from a total of more than 1600 larger oil fields from all around the planet. 

In short, our model is based on the idea that the plateau production years will be followed by 
a steeply declining oil production, as observed in the UK, Norway, Indonesia and Mexico and described 
below. We presume that this declining production profile can be used as a universal decline curve for all ageing conventional oil extraction regions of the planet. 

%  https://www.iea.org/media/executivesummaries/WEO_2013_ES_English_WEB.pdf 

\subsection{The declining oil extraction in Western Europe, Mexico and Indonesia.}

Crude oil production in Western Europe (EU plus Norway), \cite{EIAoildata}, increased steeply during the last decade of the 20th century.
The production reached a plateau like maximum of around 6.2 mbd and for about 5 years and 
production started to decline around the year 2002. 
Since 2004, and despite much higher oil prices, the extraction of oil from all Western European oil fields combined began to decline steeply. During 2014, the crude oil production was reduced to 2.87 mbd, less than half of the amount reached during the production plateau. 

Similar oil production declines are documented for Mexico and Indonesia \cite{EIAoildata}. 
During the last decade of the 20th century, the crude oil production in Indonesia reached a plateau value
of around 1.5 mbd and declined by a factor 2 during the last 14 years
to 0.79 mbd (2014).  Mexico reached a production plateau of about 3.4 mbd during the years 
2001-2006,  from were it declined to only 2.5 mbd during 2014.  

In order to simplify the comparison between these countries and regions, the yearly crude oil production
is normalised to their average plateau production values and the years are shifted by +3 years 
for Indonesia and -4 years for Mexico, such that the last plateau year is matched for all countries.
The resulting production curves are shown in Figure 1.  
While the growth period shows significant differences, when looking over  
a period of 5-10 years during the decline period, the observed decline appears to be 
rather similar for the three regions. In addition one can also observe that, during 
some smaller time intervals,  a flattening in the decline is possible. Presumably this can be 
understood due to the opening of likely less profitable smaller oil fields. 

\begin{figure}[h]
\begin{center}
\includegraphics[width=14cm]{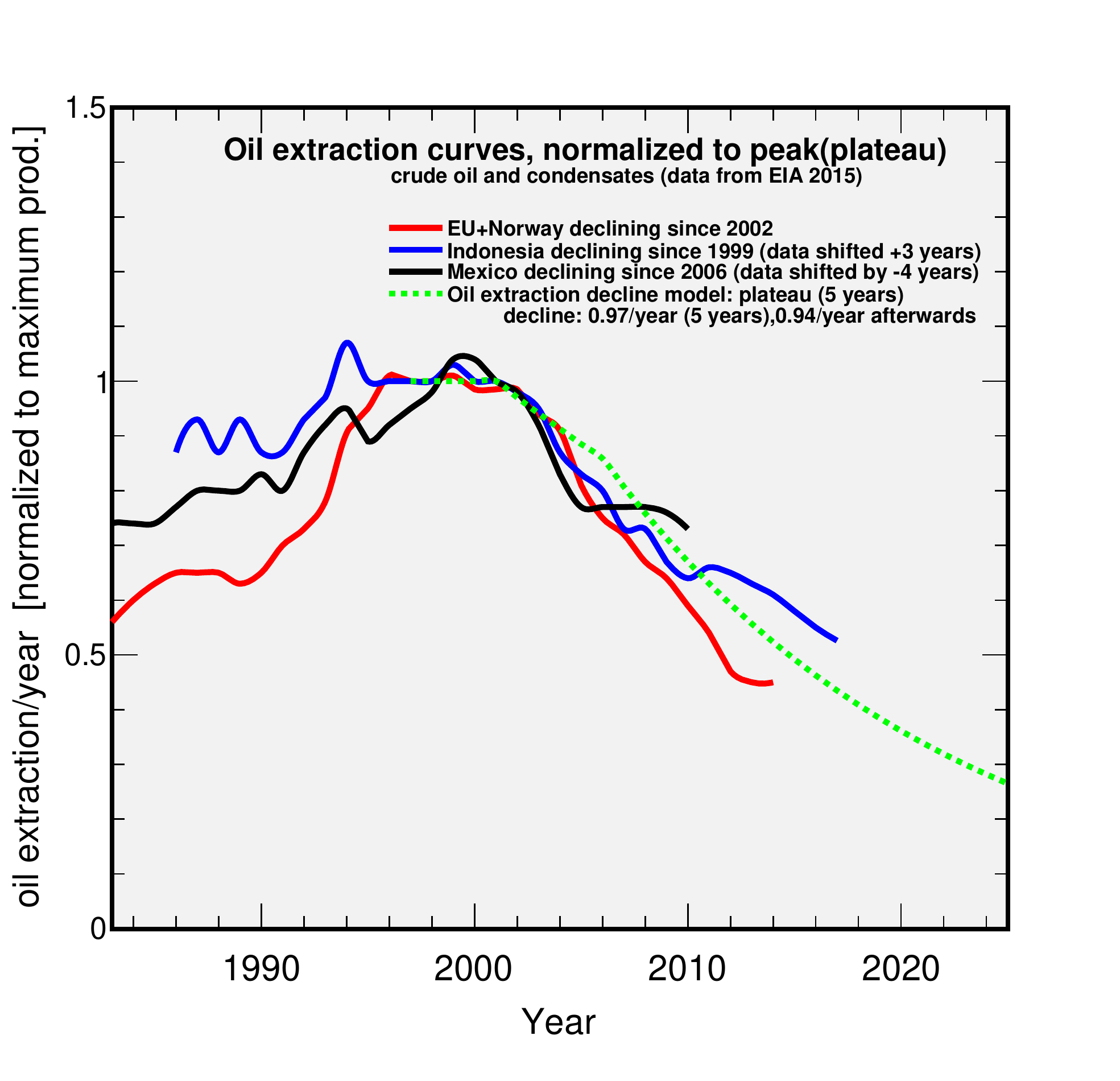}
\end{center}
\caption{Crude oil extraction curves from three declining countries and regions.
The data are normalised to each average peak production result (the average from the years close to the maximum). The peak year for Mexico and Indonesia were 2006 and 1999 respectively. 
Their data are shifted such that the "peak year" matches with the 2002 peak year from Western Europe. 
The doted curve is the prediction from the model, which assumes the plateau
ends with an annual production decline of 3\%/year for 5 years followed by a 6\%/year decline.}
\end{figure}

In order to describe the decline period, we propose a simple model which postulates that a few plateau 
years will be followed by a 5 year period with a 3\%/year average decline, followed by a 6\%/per year
decline. The resulting curve (dotted in Figure 1) provides an overall good description of the steeply declining oil extraction trends observed from Western Europe, Mexico and Indonesia. An annual production decline of 6 $\pm$ 1\% implies that, at the end of ten years, production will be only 54 $\pm$ 6\% of what it was at the beginning of those ten years. In order to use this model to estimate the future production in different regions,
the last plateau year needs to be defined and in order to minimise annual fluctuations, 
we propose to use the available average production data from the last three years as a starting point
and the production trend will be defined from the previous three years.  
Thus, according to the EIA data base \cite{EIAoildata},
the conventional oil production for Western Europe in 2012, 2013 and 2014 was 3.0 mbd, 2.83 mbd and 2.87 mbd respectively. When averaging the production during these three years one obtains the 2013* annual production of
2.9 mbd (1.06 Gb). Using the 6\%/year decline, we model the production in 2020 as 1.9 mbd (0.69 Gb).
If instead we would choose a single year between 2012 and 2014 as the starting year, the production 
would have declined to 1.9 mbd sometime between 2019 and 2021. 
Accordingly we propose to use this $\pm 1$ year as the model uncertainty if the country/region is already in the 6\%/year decline phase. If the country or region is in the plateau or -3\% decline phase the uncertainties 
must be slightly larger and for simplicity we guesstimate the uncertainty 
twice as large corresponding to $\pm 2-3$ years. 

Using the 6\% annual decline rate and the 2013* average production of 1.06 Gb, the future annual crude oil production for 2020, 2030 and 2050 is estimated as being about 0.69 Gb, 0.37 Gb and 0.11 Gb respectively. 
Assuming an exponential decline, the production will obviously never become exactly zero. However, one can use the total future production according to this decline model, and estimate the year when today's claimed reserves will be terminated.  Assuming the 6\% annual production decline, 
and starting with the 2013* annual extraction from Western Europe, Mexico and Indonesia of 1.06 Gb, 
0.93 Gb and 0.30 Gb respectively, one finds that their official crude oil reserves (end of 2013) of 
12.8 Gb, 11.1 Gb and 3.7 Gb, will all be exhausted around the year 2035. 

Assuming however that new findings or reserve revisions, due to improved extraction technology, 
will increase the 2013 reserves by about 30\%, the predictions from this decline model might become extendable up to the year 2050.  

Our simple model can therefore also be used to transform the usually quoted static reserves, the lifetime with a constant production, into a more realistic dynamic lifetime with a declining annual production.
For example, if the remaining reserves are given with a static lifetime of 10 years and the 
actual production declines by 6\%/year, the declining production can continue for about 20 years.
Alternatively, if the yet to be found reserves would increase the current oil reserves by about 30\%, 
the 6\%/year production decline period could be extended from the 20 year limit to about 35 years.
In a similar way, realistic reserves can be estimated for countries and regions which are declining by 6\%/year since some years. Assuming that those decline rates can be sustained roughly for the next 25 years, we find that an annual extraction of about 1 Gb (or 2.74 mbd) corresponds to exploitable in place reserves of about 13 Gb. Our simple model, which uses nothing more than production trends from the past few years, leads thus also to a good approximation of remaining reserves.

Alternatively, the consistency of the presented static reserves could be determined by comparing the estimated potentially possible production phase with the recent production trend in every region. For example, 
\begin{itemize}
\item static reserves of less than 15 years,  mean that the steep annual production decline of 6\% per year
has started or will be imminent;
\item static reserves between 15-20 years mean that the 5 year period with an annual production decline 
of about 3\% has began or will start now;
\item static reserves between 20-25 years mean that the plateau production can continue for at most 5 years and
\item static reserves larger than 25-30 years mean that the plateau production can either be extended for 
many more years or that there is a real possibility to increase production during the next few years.   
\end{itemize}

\section{Regional oil extraction, the next 35 years} 

Following the ideas, which lead to the simple model in the previous section,
we now use the EIA conventional crude oil production data and officially claimed reserves \cite{EIAdatabase}
from the last 5-10 years from all major producing countries, regions and continents to:
\begin{itemize}
\item Compare the actual production and the internal consumption with the reserves,
\item determine the actual production phase according to our model, separated in growth, plateau and the 
3\% or 6\% decline phase,  
\item and apply our model to estimate the future oil production up to the year 2050 in a country or region and discuss the consequences for future crude oil exports. 
\end{itemize}
Table 2 and Table 3 summarise the officially given conventional and unconventional oil  
past production and reserve data and the results from our simple model for their corresponding 
future oil production for the years 2020 and 2030. The corresponding predictions for the 
next 35 years and essentially all regions and continents are shown in Figure 2 and 3. 
%Even assuming that the remaining reserves of the Middle East OPEC countries are only around 
%50\% of the claimed reserves, today's production level can in principle be maintained at least up to 2050.   

\subsection{Western Europe}

The declining crude oil production and consumption in Western Europe (EU + Norway +Switzerland) 
has been used to develop the current model and many details were given in section 2. 
In addition to those arguments presented already, it is important to consider the fact that oil production in essentially all Western European countries is in decline and that all countries, excluding Norway, 
import a large fraction of the consumed oil. 

Following our model, we predict that the overall conventional oil production in Western Europe 
will decrease from 2.9 mbd (2014) to 1.9 mbd (2020), 1.0 mbd (2030), 0.55 (2040) and less than 
0.3 mbd around the year 2050. 

As we are trying to model the future production and consumption in each region, 
the crude oil production in Norway, the only exporting country in Western Europe, is of particular relevance to its oil-importing neighbours. Conventional oil production in Norway has declined from 2.95 mbd (2004) to 1.57 mbd (2014), and more than 80\% of this oil was exported in 2014 to the neighbouring EU countries.

{\tiny
\begin{table}[h]
%\begin{center}
\begin{tabular}{|c|c|c|c|c|}
\hline
Region & reserves  & production & production     &modelled production   \\
            & [Gb]         & [mbd year]    & phase 2015  &     [mbd year]  \\
            & 2014   & 2010* / 2013*   &  [\%/year]     & 2020 / 2030   \\ 
             \hline
Norway      &     6     & 2.12  /   1.57                &    -6\%                             &  1.02 / 0.55          \\
EU             &     6     & 1.77 /    1.33                &    -6\%                             &  0.86 / 0.46           \\
\hline
Western Russia      &    $\approx$ 60    &       9.4 / 9.3             & -3\%                 &  7.5 / 4.0     \\
Far Eastern Russia&    $\approx$ 20    &       0.3 / 0.7   & $\geq$ +10\%             &  1.6 / 1.4     \\
Azerbaijan         &              7     &       1.0 / 0.88  &  -3\%                &    0.63 / 0.34         \\
Kazakhstan       &              30   &        1.51 / 1.57 & $\approx$ +1\%           &   1.8 / 1.4              \\
\hline
China      &                     24   &    4.0  / 4.15        & plateau          &  3.6 / 1.9               \\
India        &                     5.5    &   0.73  / 0.78    & plateau         &   0.78 / 0.5               \\
other Asia/Oceania &      15.5     &  3.01  / 2.76   & -3\%              &  1.9 / 1.0             \\
\hline
Libya*    &                 48       &   1.3 / 0.9  &  plateau$^{+}$&  1.5 / 1.5                \\
Algeria   &                  12        &  1.56 / 1.47     &  -6\%        &   1.0 / 0.53                                      \\
Angola   &                   9            &  1.84 / 1.78     &  -3\%        & 1.35 / 0.73                                 \\
Nigeria   &                  37        &  2.41 / 2.44     & plateau   & 2.5 / 1.6                                    \\
other Africa &             22           &  2.23 / 1.9     & plateau$^{+}$   &  2 / 1.2                                 \\
\hline
Saudi Arabia&             268       &  8.9  / 9.75   & plateau$^{+}$        &  10 $\pm$ 1                                    \\
other OPEC Persian Golf &  524 & 12.9 /  13.4  & plateau$^{+}$        & 14 $\pm$ 1.5                          \\
\hline     
Brazil &                   15       & 2.04 / 2.11            &  $\approx$ 1\%    &  1.8 / 1.0                 \\
Venezuela* &            80     & 2.47 / 2.5            &    plateau$^{+}$    &  2.5 / 1.6+0.9             \\
other South/Central Am.& 13  & 2.2 / 2.33       & ``plateau"             &  2.1 / 1.35                   \\
\hline     
USA (convent.) &        23       & 4.6 / 4.3           &  -3\%      &  3.7 / 2.0         \\
USA (tight) &       14*           &  0.9 / 3.2            &  $\geq$ 10\% &  4.5 / 4.5  \\      
Canada(convent.) &   6*                  &  1.23 / 1.12          & -3\%       &  0.85 / 0.5          \\
sands+tight &  167  & 1.5 / 2.2            &  $\geq$ 10\%   &   3  / 3 \\
Mexico &   10   &  2.62 / 2.54           & -6\%         &  1.6  / 0.9   \\
%World (convent.) & 71.7 / 71.3          & -               &   63 /  46 \\
\hline
\end{tabular} %\vspace{0.1cm}
\caption{Conventional crude oil reserves and production numbers from the EIA (2014) \cite{EIAoildata}.
The average oil and condensates production from the years* 2009-2011 and 2012-2014 are separated 
for conventional and the tight oil and tar sand production. The production forecast for 2020 and 2030 is the result obtained from our model. The plateau$^{+}$ stands for regions where the past years oil production was limited by political and economic conditions. For Venezuela we assume that the decline will be compensated by some heavy oil production
starting during the next decade. 
} 
%\end{center}
\end{table}
}

Following our model\footnote{The future production from the Johan Sverdrup oil field, discovered in 
2011, might lead to an additional production of perhaps 0.55 mbd from 2019 onwards \cite{Campbell16}. However, 
this potential production is currently excluded in our estimate.}, the conventional oil production in Norway will fall to 1 mbd by 2020.   
Assuming that the internal oil consumption will stay at today's level and that the production from the new large Sverdrup oil field will not be happening according to the plans, the exports to the EU will decline accordingly. 
During 2014, 41\% of the exported oil was sent to the UK, 27\% to the Netherlands and 12\% to Germany \cite{Norwayexports}. 
So it is especially important for the UK and the Netherlands to find other oil-exporting sources as quickly as they can. As will become clear in the next subsection, neither Russia nor the other 
oil exporting countries from the FSU will be able to replace the missing million barrels of oil during the next 
decades. 

\subsection{RUSSIA and other FSU countries}
The three Former Soviet Union countries with substantial oil production and reserves are 
Russia, Kazakhstan and Azerbaijan. During 2013/14, about 6 mbd of crude oil were exported from these countries, mainly from and through Russia to Western Europe \cite{RussiaFSUEIABP}.

During the last years, the total production in Russia reached a plateau of 10 mbd (2014). However,  
it is important to realise that about 90\% of this oil is produced from relatively large oil fields in the 
western parts of Russia and of Western Siberia. Most of these fields are in little populated regions of Western Siberia and are exploited since decades. From there the oil is transported through a vast pipeline system to the European part of Russia and from there to Western Europe \cite{RussiaFSUEIABP}, \cite{Russiaoxford}.

The other 10\% of the produced crude oil comes from relatively new explorations. Most of the new 
explorations are in Eastern Siberia, and this oil will be transported through existing and planned pipelines to China and other Asian countries. Subtracting the growing production in Eastern Siberia from Russia's total production, it seems that the first decline phase (3\%/year) in the Western parts of Russia has began already or is imminent. This seems to be confirmed by the latest statements from the Russian oil company Lukoil and also from a recent 
IEA report \cite{RussiaIEA}.    

Consequently, the year 2015 will be assumed here to be the beginning of the 3\%/year decline phase in Western Russia and Western Siberia, followed by the 6\%/year decline from 2020 onwards.
The oil production from Eastern Siberia, if production goes more or less
according to plans, should rise from about 1 mbd in 2014 to 1.6 mbd by 2020 \cite{Russiaoxford}. We assume that production will stay at this plateau for 5 years to 2025 and will start to decline by 3\%/year to 2030 and 6\%/year afterwards. 
Combining our model with these assumptions, we find that the production in 2020 will be about 7.5 mbd from Western Russia and 1.6 mbd in Eastern Siberia. Calculated accordingly, the total Russian crude production in 2030 will be about 5.4 mbd. 

The official Russian reserves increased according to the EIA data base \cite{EIAoildata} from 
60 Gb (2012) to 80 Gb (2013). Assuming that this increase came mainly from the inclusion of the new 
findings in Eastern Siberia, the Western and far Eastern Russia reserves at the end of 2013 
are guessed as 60 Gb and 20 Gb respectively.
Following the modelled oil extraction profile, about 6 Gb in Western Russia and 5.5 Gb in Eastern Siberia 
would remain after 2050. 

The crude production in Azerbaijan rose quickly from 0.5 mbd to 1 mbd between 2005 and 2011, 
but declined by 15\% during the last 3 years to 0.85 mbd in 2014. We assume that the 3\% decline phase is ongoing and will continue to 2016. Afterwards, the production will decline by 6\%/year and the production in 2020 and 2030 is modelled to be 0.63 mbd and 0.34 mbd respectively. Accordingly, the current reserves, given as 7 Gb, would be reduced to about 2 Gb by the year 2050. 

The official reserves of the landlocked country Kazakhstan are given as 30 Gb. Between 2010 and 2014
crude oil production grew by about 7\%. New and planned oil explorations and pipelines are all found in the eastern parts and it thus seems logical to assume that the additional oil will all be exported to China.
For the older oil fields, mostly connected through pipelines to the Russian infrastructure, we assume that 
they will start the decline during the next 5 years. Combining this with some continued growth in the eastern parts, a plateau production at around 1.8 mbd will be reached between 2018 and 2023. After that and applying our model, the production will decline by 3\%/year to 2028 and 6\%/year afterwards.  We thus estimate the production in 2020 to be 1.8 mbd from 
where it will decline to 1.4 mbd by 2030. 
According to this modelled production profile, the remaining reserves around 2050 would be about 15 Gb. 
One might conclude that today's reserves of Kazakhstan are either overestimated, or that 
additional production and export capacity of up to 2 mbd might be developed during the next decades. Considering the geographical location and the growing importance of China in this region, it seems most likely that any new and expensive pipeline structures would deliver this additional oil to China.
 
Combining the internal oil consumption in the FSU countries, given as 4.4 mbd during 2014,
we assume that the internal consumption increase observed during the past years might continue for a few more years to at least to 4.5 mbd in the year 2020. Given that increase in the FSU's consumption and the decrease in its production, oil available for export to Western Europe will decline from 6 mbd in the year 2014 to about 4 mbd (2020) and will essentially be terminated around the year 2030.

\begin{figure}[h]
\begin{center}
\includegraphics[width=14cm]{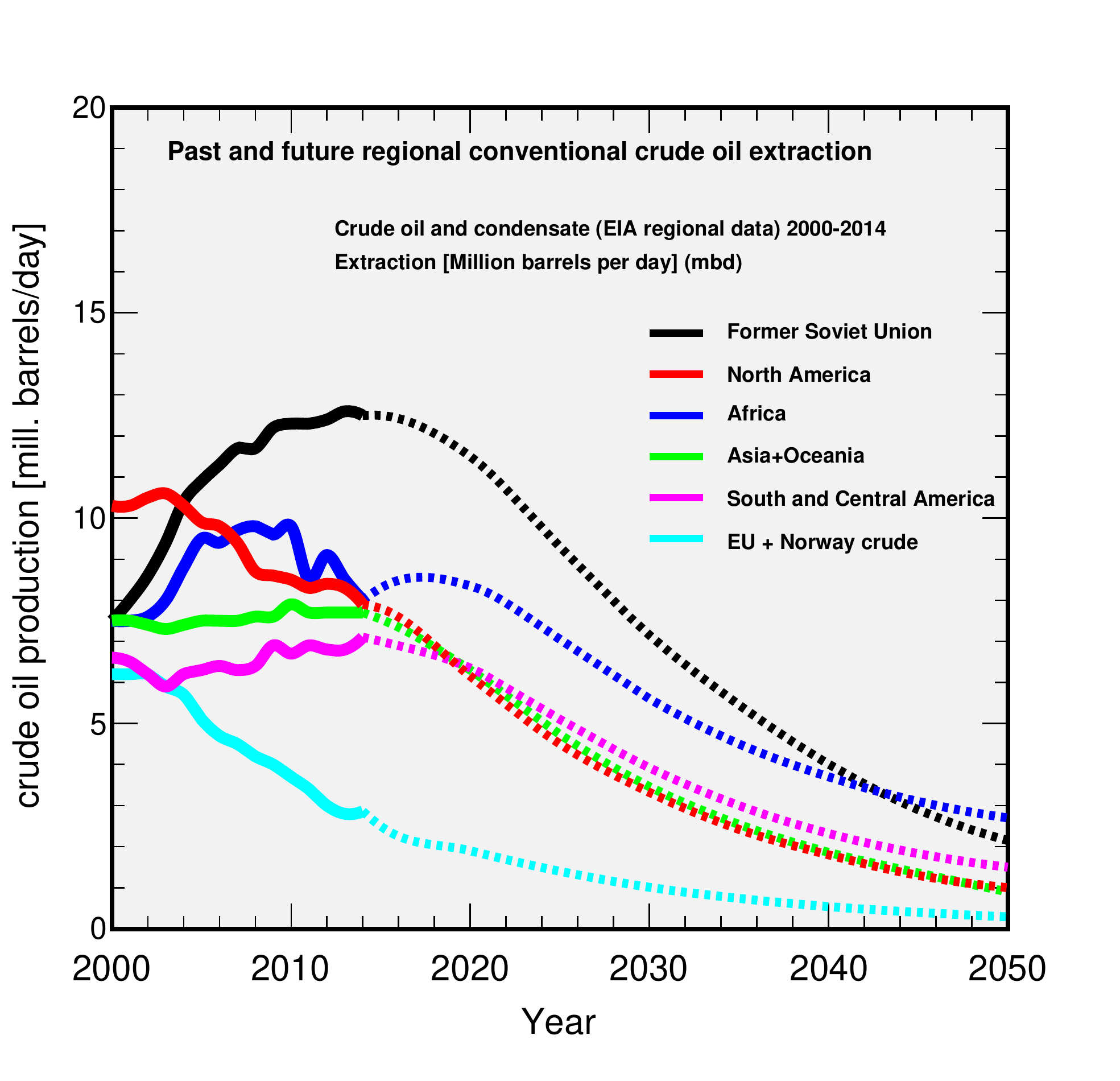}
\end{center}
\caption{\small{The conventional crude oil production data from 2000-2014 (full lines) and the modelled future 
production up to the year 2050 (dotted lines), for all major oil producing regions and continents, outside the Middle East. 
%The oil production from the Middle East OPEC countries increased during the last 5-10 
%years to about 24 mbd (9 Gb per year). Without ``war like interruptions" and assuming that the claimed %remaining reserves of these countries are at least half as large as claimed, we expect that their current %production can be maintained or even be increased during the next decades.
}}
%\end{center}
\end{figure}

\subsection{Asia/Oceania} 
The remaining crude oil reserves in the Asia and Oceania region, where almost half of the 
world population lives, are relatively small and are given as 45 Gb \cite{EIAoildata} at the beginning of 2014.
The overall production stayed roughly constant during the last 10 years and about 2.5 Gb of crude oil was produced during the year 2014. During the same period, the annual consumption in this region increased by about 30\% from 7.5 Gb to about 10 Gb (27 mbd). Most of this imported oil originates from the Middle East OPEC countries and is transported by large ships that carry up to 3 mb.

Slightly more than half of the estimated remaining reserves and of the current production in this area is  associated with China.  The crude oil production in China grew during the first decade of the 21st century, 
but appears to have reached the plateau production phase. We model the future production 
for China at a plateau value of 4.2 mbd until 2015, followed by 
5 years with 3\%/year decline and by 6\%/year afterwards. Production would decrease accordingly to about 3.6 mbd 
in 2020 and 1.9 mbd in 2030. Following this extraction model, China's official reserves, 24 Gb, would be terminated around the year 2050. 
 
India's reserves at the beginning of 2014 are given as 5.5 Gb and production over the past decade was rather stable at about 0.7-0.8 mbd. This is still far below the relatively small internal consumption of 3.7 mbd (1.35 Gb/year). 
We assume that this production plateau can be maintained to 2020, and will decline by 3\%/year to 2025 and by 6\%/year afterwards. According to the modelled future production, India's reserves will be 
exhausted around the year 2050.  

Combining the other Asian and Oceanic oil producing countries, mainly Indonesia, Malaysia, Australia, Thailand and Vietnam, one observes that their production since 2009 declined by about 13\% and thus almost by 3\%/year. We assume that the 6\% decline period starts in 2016. Accordingly, the production will decline from 2.76 mbd (2013) to 1.9 mbd and 1.0 mbd by 2020 and 2030 respectively. With reserves of about 15.5 Gb (end of 2013) and our modelled production, the reserves are expected to be terminated between 2045 and 2050.

\subsection{Africa} 
Oil production perspectives on this huge continent, with a population of 
about 1 billion people, can be comfortably divided into Northern Africa and the Sub-Saharan Africa.  
The total oil reserves of Africa at the beginning of 2014, given by the EIA, are 127 Gb. 
About 83\% of these oil reserves are found in four countries: Libya (48 Gb), Nigeria (37 Gb), Angola (9 Gb) and 
Algeria (12 Gb). About 76\% of the African oil in 2014 was produced by these four countries. 
With an overall production of 8 mbd and an almost negligible per capita oil consumption of 0.6 litre per day, more than 50\% of the produced oil in 2014 was exported to Western Europe (2.8 mbd), China (1.3 mbd), 
India (0.6 mbd) and North-America (0.6 mbd) \cite{BPstat15}. \\

{\bf Northern Africa}\\

The oil production in Libya, with its small population, is heavily affected by the ``aftershocks" of the recent 
war against this country. The 2014 production was 0.5 mbd, about 1/3 of the production before the war. Even if the presumably huge and high quality crude reserves of  48 Gb would be largely exaggerated and without further wars, it should be possible to increase the oil production at least to the prewar level of 1.5 mbd (0.55 Gb/year)
and maintain it at that level for many decades. We assume that around the years 2020 and in the following decades the production will reach the prewar production of around 1.5 mbd. 

Algeria, a country with almost 40 million people and with a very young population, had managed to 
increase the oil production between the year 2000 and 2008 by about 40\% to 1.7 mbd. During the following years the production declined by almost 20\% to 1.4 mbd in 2014. 
At least since 2004, the reserves are given as 12 Gb and were not changed according to the yearly extraction of about 0.5 Gb. The remaining reserves can thus be guessed to be at least 5 Gb smaller than officially given. The decline during the last years corresponds to an average of about 3\%/year. Accordingly we expect that the yearly 6\%/year decline phase starts during 2015. The resulting annual oil production is thus modelled to be about 1 mbd (0.36 Gb/year) and 0.5 mbd in 2020 and 2030 respectively. Depending on the actual true reserves, the declining production might last until 2050. 
Following this production decline and assuming that the actual internal consumption, about 0.4 mbd in 2014, will stay at the current level as long as possible, oil exports from this country, mainly to France,  will end around the year 2030. 

The third North African country, with some small oil production is Egypt. 
The production in 2014 was about 0.5 mbd, slightly smaller than in 2009 when almost 0.6 mbd were produced.  As this decline is probably more related to the unstable political situation in the country, 
we expect that the current production can be maintained at the current level of 0.5 mbd until 2020.  
We assume that the 3\%/year decline phase will
start in 2021 followed by the 6\%/year decline phase in 2026. The resulting modelled oil production 
will be about 0.5 mbd in 2020 and 0.32 mbd in 2030. The total production up to 2050 would add up to about 4 Gb. This matches with the reserve estimate given as 3.6 (BP) and 4.4 Gb (EIA).

Considering the long term historical cultural relations between all countries in Northern African, 
and especially the common border between the highly populated Egypt and the small population of 
Libya, one can imagine that the Libyan oil will eventually be directly shared with the  
people in Egypt and possibly also with the people in the other North African countries.\\

{\bf Sub-Saharian Africa}\\

With 2.4 mbd, Nigeria is currently the largest oil producer in Africa and almost 90\% of this oil is exported.  
With several new production areas under preparation, the production could 
potentially be increased during the next decade to more than 3 mbd. However, civil war and other political instabilities within the country resulted in large production fluctuations and major delays of new projects.
If the remaining reserves in 2015 are indeed 37 Gb, and if the civil-war-like problems for the production 
can be avoided during the next decades, it might be possible to increase the production 
to perhaps 3.5 mbd and remain at this level for several decades.

As in many other countries, it is unclear if the currently reported reserve number represents
the actual reserves or the original in place reserves. Assuming that the reported 37 Gb refers to original in place reserves, the total production between 1980 and 2014, about 25 Gb, needs to be taken into account for the future production estimates. Under this extreme scenario, 
the remaining reserves would be more like 12 Gb, and the current production could only 
be maintained for another decade. 
In absence of better data, we assume that the current plateau like production of about 2.5 mbd will be maintained until 2020. Afterwards we assume that the production will start to decline 
by 3\%/year and 6\%/year from 2025 onwards. We thus model a total extraction from 2015 to 2050
of about 17 Gb. 

The other important oil producer and exporting country is Angola. The production in 2014 was 1.74 mbd,
about 5\% smaller than the average production between 2009-2013. The years   
2014-2018 are thus assumed to correspond to the 3\%/year decline period. Accordingly, the modelled 2020 and 2030 production will be about 1.4 mbd and 0.7 mbd respectively. The total modelled oil production until 2050 is thus predicted to be about 10 Gb, which matches the reserve numbers given by BP of 12.7 Gb
and the EIA of 9 Gb. 

The production from all the other smaller producing Sub-Saharian African countries in 2014 was 
1.4 mbd. This significant decrease from the 1.7 mbd produced in 2010 might be explained by the very 
unstable political and economic situation in many African countries. Like for Egypt, we assume that the 
current production can be kept stable until 2020, and that the 3\%/year decline will start in 2021. We thus model the total production to be 1.5 mbd and 0.9 mbd in 2020 and 2030 respectively. Their combined total production to 2050 corresponds to about 12 Gb, which is slightly smaller than the totalled estimated reserves of 15.2 (BP) and 16.6 (EIA).
    
\subsection{South and Central America} 

Overall, only Brazil, Venezuela and Ecuador are known to have considerable oil reserves.
However, a large fraction of these reserves, are very difficult to exploit. Examples are the unconventional oil sands in Venezuela (about 220 Gb of the 298 Gb), the recently found very deep sea oil in Brazil (about 10 Gb of the total 15 Gb) and the oil in the biodiversity rich and partially protected Amazon region of Ecuador (total 8.8 Gb). The reserves of all other countries in South and Central America combined are given as 7.2 Gb.    

During the last 5 years, the crude oil production in South and Central America increased by 
7\% from 6.7 mbd (2010) to 7.1 mbd (2014). The consumption of all liquids during this period increased 
by 14\% to about 7.1 mbd in 2014.
Most of this production increase came from Brazil (from 2.06 to 2.26 mbd) and Colombia (0.79 to 0.99 mbd). Brazil, despite its production increase, remains a large importer with an annual average of about 1 mbd. Due to the relatively low internal per capita oil consumption, Venezuela, Columbia and Ecuador 
consumed internally only a small fraction of the produced oil. The 2014 exports are reported as 
1.9 mbd, 0.7 mbd and 0.3 mbd respectively.

Focusing first on Venezuela, it is important to notice that today's oil production, about 
2.5 mbd, is 30\% lower than during the peak years around 1970 and between 1995-2000.
The supposed huge reserves of this OPEC country are known to have very low quality and the numbers appear to be very political as large upward changes were reported during the last decades. 
Between 1980 to 2010 the claimed conventional reserves increased from 18 Gb to 59 Gb (1990), 73 Gb (2000) and 99 Gb (2010). The potential oil sand reserves of 220 Gb were added to the total reserves during the next years and the official reserves increased further to 211 Gb (2011) and to 298 Gb (2013). 
The exploration prospects of these unconventional reserves will be discussed in section 3.8. 

The more conventional reserves are claimed to be about 80 Gb, almost as large as the reserves associated with Russia. Even if these 80 Gb are considered to represent the original in place oil reserves around the year 1980, about 45 Gb reserves should remain as about 35 Gb have been extracted during the last 35 years.
Even such reduced reserves would be roughly twice as large as the ones in the USA and China. One would thus expect that the daily production could perhaps be increased to 3.5-4 mbd and remain at this level for some decades. In absence of convincing reserve data and following the plateau like production during the last decade, we assume that the current annual oil production of Venezuela, about 2.5 mbd (0.91 Gb/year), can potentially be maintained for several decades.
If this production would be maintained to 2050, a total of about 30 Gb would be extracted which is still smaller than the claimed remaining or original in place reserves. Perhaps it is more realistic 
to assume that the remaining reserves are much smaller than claimed and that the 
production will start to decline by about 3\%/year during the next 5-10 years. We would thus expect 
that the production will start to decline during the next decade and will decline from 2020 
onwards by 3\%/year and will reach about 1.6 mbd (2030). Adding a possible production 
from the huge oil sand reserves (see section 3.8) we assume that the overall oil production 
might approximatively remain untill 2050 at today's level of 2.5 mbd. 

The situation for Brazil looks even more difficult to predict, as the access to about 2/3 of the claimed reserves appears to be technologically extremely difficult. Since it now appears unlikely that the country's deep-sea oil Þeld exploration will begin any time soon, \cite{Petrobas}, we assume that the plateau value has been reached already and that the 3\%/year decline will start in 2016. Accordingly we model a production of 1.8 mbd by 2020 and 1 mbd in 2030. This production decline scenario results in a total production of about 7.5 Gb by 2025 and would still require that some substantial fraction of the deep sea oil reserves, about 10 Gb, can be successfully exploited during the next decade. 
 
The future oil production in Ecuador, with official reserves of 8 Gb, 
depends critically on biodiversity protection measures which might be taken in the Amazon region. 
As it seems unlikely that any clear decision will be made during 
the few next years,  we assume that the current annual production of about 0.5 mbd (0.2 Gb/year) can be maintained during the next decades.  

Combining all other producing countries in South and Central America, a rather constant 
crude oil production of about 1.8 mbd was observed during the last 5 years.
Accordingly, we expect that the 3\%/year production decline period will begin in 2016 and followed by 
6\%/year from 2021 onwards. 
The modelled production is estimated to be 1.57 mbd (2020) and 0.85 mbd (2030). The resulting total production up to 2030 would be about 7.5 Gb, which exceeds the current claimed reserves of 5.2 Gb.  
  
Combining the above production numbers, we expect that the total conventional oil production within South and Central America will decrease from the current level of about 7 mbd to 6.3 mbd around 2020 and 4.8 mbd (2030).
Given the uncertain prospects of Brazil's deep sea oil fields, and given the lack of clarity regarding the ratio of conventional to unconventional oil reserves in Venezuela, the estimated future maximum oil production in South and Central America is probably overestimated.

\subsection{North America}

In this section we discuss only conventional crude oil production, current and future, in the USA, Canada and Mexico.
The situation and prospects with unconventional oil and other liquids,
which increased rapidly during the past few years, especially and dominantly 
in the USA and Canada, will be presented in detail in section 3.8. 

Oil production in Mexico started to decline in 2006 and has reached the average decline rate 
of about 6\%/year during the last few years. As the situation in this country was used 
to develop the future production model, we refer for more details to section 2.
However, it is important to note that Mexico produced about 2.5 mbd crude oil plus 0.3 mbd of other liquids
and consumed about 2 mbd oil-equivalents per year (about 1000 liter/capita). During past years, most of the oil exports were sent to the USA. With a roughly constant internal consumption since 2006, the exported oil has decreased steeply from about 1.7 mbd in 2006 to 0.7 mbd in 2014.
Assuming that internal consumption during the next few years remains roughly at today's level, the current 
6\%/year decline leads to the conclusion that by the year 2020 Mexico will essentially stop being an oil exporting country.

After many years of declining crude oil production in the USA,  the situation has changed dramatically during the last few years. Crude oil and condensate production decreased steadily from 
8.6 mbd in 1980 to 7.4 mbd, 5.8 mbd and 5.0 mbd in the years 1990, 2000 and 2008. 
Since 2008 however the trend has changed, and production in 2014 is reported to be 8.65 mbd, which is essentially as high as 35 years ago. This production increase was possible due to the new tight (shale) oil technology, which allowed exploitation of unconventional and relatively large oil resources.

In order to model future production of conventional crude oil, the tight oil production has to be subtracted from total liquids production, just as production of other oil liquids must be.
According to the EIA reference, \cite{EIApasttightoil}, one finds that the tight oil production in the USA went up from about 0.1-0.2 mbd between the years 2000 and 2007 to 0.5 in 2008 and from there to 
2.2 mbd, 3.2 mbd and 4.2 mbd in 2012, 2013 and 2014 respectively. Details about the future possible shale oil production and other unconventional oil resources will be discussed in section 3.8. 

The conventional crude oil production in the USA has thus continued to decline by about 1.5\% to 2\%/year from 8.6 mbd (1980) to 5.7 mbd (2000) to 5.1mbd (2005) to 4.7 mbd (2010) and to 4.5 mbd in 2014. This observed decline rate during the last decades is smaller than what would be expected from our decline model. But this smaller decline rate can be easily explained by the contributions from the recently opened off-shore oil fields in the Gulf of Mexico along with increased production in Alaska. If more new fields were to be opened it would obviously reduce the overall production decline rate.   
The officially declared remaining crude oil reserves in the USA have increased from 23 Gb (2008), before the tight oil reserves were included, to 37 Gb in 2014. It appears logical to attribute about 23 Gb for the conventional reserves and 14 Gb to the exploitable tight oil reserves. 

Accordingly, one observes that the current conventional oil reserves of the USA and their 2014 production data match roughly with the 2014 numbers from China, presented in section 3.3. 
Without detailed information about a possible extraction from additional off-shore or arctic oil deposits, we assume that the already declining conventional crude oil production in the USA will start the 3\%/year and 6\%/year decline phases in 2016 and 2021, respectively. The conventional USA oil production is thus modelled as 3.7 mbd (2020) and 2 mbd (2030). 
The modelled annual conventional oil production will exceed the official reserves around the year 2040.   

The conventional plus unconventional oil production in Canada is reported as 3.3 mbd and 3.6 mbd in 2013 and 2014 respectively \cite{EIAoildata}. Conventional crude oil production in Canada is obtained by subtracting, from total production, the tar-sands production (2.0 mbd (2013) and 2.2 mbd (2014)) and the tight oil production (0.3 mbd (2013) and 0.4 mbd (2014) \cite{Canadaoil}).
It follows that the conventional oil production declined from an average of 1.23 mbd (2009-2011) to 1.12 mbd (2012-2014) respectively. In absence of more detailed data, we assume that the 5 year 3\%/year decline period goes from 2013 to 2018 followed by the 6\%/year decline. The corresponding oil reserves are not given independently, but using BP's, \cite{BPstat15} estimated 167 Gb of ``tar sand" reserves", one finds the conventional crude oil reserves are about 6 Gb. The total production would exceed these 6 Gb a few years after 2050. 

Combining the three countries and following our model, we expect that the maximal possible conventional oil production will decline 
from 7.9 mbd (2014) to 6.2 mbd (2020) and 3.4 mbd (2030). Adding a roughly constant production of unconventional oil 
from tar sands in Canada (2.5 mbd) and a total tight oil production of about 4.5 mbd for the years 2020 and 2030 respectively (for details see section 3.8), the combined conventional and unconventional crude oil production from North America is estimated to decline from about 14.7 mbd (2014) to about 13 mbd (2020) and 10 mbd (2030).

\subsection{The Persian Golf OPEC countries}

The combined oil production of the six Persian Golf OPEC countries - Saudi Arabia, Iraq, Iran, Kuwait, UAE and Qatar - amounted to 23.4 mbd (about 8.5 Gb/year) in 2014.
Their combined official crude oil reserves, either in place or originally, are given by BP and EIA as about 800 Gb, 
or as roughly 61\% of the global total conventional oil reserves. 
Even if their remaining reserves are exaggerated, and even if the roughly 230 Gb produced during the last 35 years should be subtracted from the estimated 800 Gb, their combined remaining reserves are still huge.

In fact, most of these Middle East OPEC countries might have some real and some theoretical spare oil 
producing capacity. Considering that it is far easier and thus cheaper for those countries to 
extract their oil than it is for any other region, there seems currently no reason to 
even consider large investments to increase the production and export capacities.
Assuming that wars, like the one in 2003 when the USA and its allies attacked Iraq, can be prevented, 
we expect that these six countries can keep their combined production within roughly $\pm$ 10\% of today's level of 24 $\pm$ 2.5 mbd at least to the year 2050.

\subsection{Production of unconventional oil and oil-equivalent liquids}

The production of unconventional oil and oil-equivalent liquids increased steeply during the last decade, corresponding today to  roughly 20\% of the worlds total liquid energy consumption. 
The largest contribution to unconventional oil-equivalent liquids comes from "Natural Gas Liquids" (NGLs, \cite{NGL}) which are a byproduct of natural gas production.

%The largest contribution came from the 
%NGL's (natural gas liquids,\cite{NGL}) which contributed globally about 10.1 mbd during 2014. 
%  http://www.eia.gov/todayinenergy/detail.cfm?id=5930 
%The 2014 production of tight oil (or shale oil), dominantly from the USA and Canada, 
%were 4.2 mbd and 0.4 mbd respectively.  In addition, Canada produced about 2.2 mbd 
%from the oil tar sands in Canada. Globally another 2.7 mbd 
%came from other oil-equivalent liquids (biomass and coal to liquids).  \\
%
It is generally accepted that the extraction of unconventional oil is technically very difficult, less energy efficient 
and more polluting than the extraction of conventional crude oil. In addition the comparison is further complicated by slightly different usages of these different oil-equivalents liquids. 
An easy and commonly used way to estimate these differences is to compare the monetary cost of extracting conventional and unconventional oil.

However, this approach is certainly limited as the environmental costs and the political costs to extract and transport the oil safely to the consumer are difficult to calculate. 
It is for example difficult to compare the monetary costs of non-conventional oil produced in the USA with the costs of the conventional oil produced in Eastern Siberia that must be transported over very long distances. The cost of pipelines, or similar transport infrastructure, for conventional oil that must be transported long distances might thus favour more local and more expensive unconventional oil production.

The monetary oil costs for producers and consumers are even more difficult to estimate when one thinks about the problems to obtain the oil from unstable countries like Iraq. And monetary costs are even more difficult to estimate if one is dealing with a country where all or parts of which are war zones or potential war zones.

A fair comparison should thus also include also the ``secure" production and transport cost from the production site to the refinery or buyer. Especially the absence of pipelines or similar transport 
infrastructure for the usage of far away produced conventional oil, might thus favour the more local and more expensive unconventional oil production and usage. 

Consequently, some might argue that the costs of local tight oil and tar-sand oil production are lower than the potential costs of oil imports from unstable regions like the Middle East. Keeping these problems with unconventional oil and oil-equivalent liquids in mind, the maximum possible production perspectives, following the EIA and similar economical estimates, for the next decades are guesstimated in the following.\\

 {\bf Production of NGL's}\\
   
The largest contribution to unconventional oil-equivalent liquids comes from "Natural Gas Liquids" (NGLs) which are a byproduct of natural gas production.
   
%The largest contribution to unconventional oil-equivalent liquids comes from the NGL's, which 
%can perhaps be partially understood as a possible byproduct of the natural gas production.
% 
NGL production increased globally from 6.4 mbd in the year 2000 to 10.1 mbd in 2014 \cite{EIAoildata}. SigniÞcant production, with more than 1 mbd, came from the USA and Saudi Arabia with a production of 3 mbd and 1.8 mbd respectively. According to the estimates from the EIA, \cite{EIAngl}, NGL production in the USA might further increase by 30\% to a plateau of about 4 mbd between 2020 and 2040. In absence of more data, we assume that NGL production in different regions will increase by a similar amount and will globally grow to a plateau value of about 13 mbd oil-equivalent between 2020 to 2030 and remain at this level until 2050.\\

{\bf Tight oil production}\\
   
The extraction of unconventional oil from shale deposits in the USA and Canada increased considerably during the last 5 years. Tight (shale) oil production in the USA and Canada increased respectively from only 0.2 and 0.0 mbd in 2007 to 4.2 and 0.4 mbd in 2014. The EIA, in its 2015 Energy Outlook, foresees USA tight oil production reaching a maximum of 5.6 mbd in 2020. From there it is expected to decline to 4.8 mbd in 2030 and 4.3 mbd in 2040.

According to a different study, \cite{futuretight}, USA tight oil production might increase only a little more and reach a plateau of around 5 mbd between 2020-2030 from where it will decline to 4 mbd around 2040. The authors of this study estimated that global tight oil production will be at most 7.5 mbd between 2030 to 2035 and 6.5 mbd around 2040 with about 2.5 mbd coming mainly from Argentina, Canada and Russia.

{\tiny
\begin{table}[t]
\begin{center}
\begin{tabular}{|c|c|c|c|c|}
\hline
Region & production & production & production     &modelled production   \\
            & [mbd]         & [mbd/year]    & phase 2015  &     [mbd]  \\
            & 2010   &      2014   &              & 2020 and 2030   \\ 
             \hline
\hline 
World (oil-equivalent)      & 85.7    & 90.7      & $\approx$ 1\%   &  $\leq$ 90     /  77  \\
\hline
World (convent. oil)   &  72.4           &  71   &  plateau     &  $\leq$ 66  / 50 \\  
World (NGL)              &  8.9           &  10.1           & $\approx.$ 3\%   &   13  / 13 \\
World (tight+tar sands)   &  2.3           &  6.8   &  $\geq$ 10\%    &  $\leq$ 8  / 11 \\  
World (other liquids)  &  2.2           &  2.7         &  $\approx$ 5\%   &  $\leq$ 3  / 3 \\
\hline
USA (NGL)               &  2.1           &   3.0           & $\geq$ 10\%  &   4  / 4      \\
Middle East (NGL)   &  3.0            &  3.1            & plateau  &    4  / 4    \\
Other (NGL)            &   3.8           &  4.0            &  $\approx$ 2\% &    5  / 5     \\
\hline 
USA (tight)               &   0.8           &  4.2         &  $\geq$ 10\%  &  5.0  / 5.0 \\   
Canada (tight)          &   0             &   0.4        &  $\geq$ 10\% &   0.5 / 0.5 \\   
Other (tight)          &   0             &   0               &  ``starting" &   2.0 / 2.0 \\   
\hline 
Canada (tar sands)     &  1.5     & 2.2             &  $\geq$ 10\%   &   2.5  / 2.5 \\
Venezuela (oil sands)  &   0     &  0                 &  ``starting" &   0.  / 1. \\
\hline 
\end{tabular} %\vspace{0.1cm}
\caption{Production data, \cite{EIAoildata}, for 2010 and 2014 and our guesstimated global and regional upper production limits for 2020 and 2030 for the various types of oil and oil-equivalents. Total global production uncertainties are dominated by 
the $\pm$ 2.5 mbd attributed to the future production in the Persian Golf region. 
The so called refinery gains are not included in these numbers, but are estimated by the EIA as being globally about 2.5 mbd for 2014.} 

\end{center}
\end{table}
}

For our overall forecast model, we use the above numbers as the Òupper production limitÓ and assume that tight oil production in the USA and elsewhere will increase to about 5 mbd around 2020 and about 7 mbd after 2025. However, it seems that the latest EIA production data indicate that tight oil production has peaked during the summer 2015 and that production will decline during the next 2 years \cite{USAoil1617}. If this production decline is indeed seen during this short period, we believe actual production during the next few years will very likely be signiÞcantly less than our assumed upper production limit.
\\

{\bf Oil production from tar sands, biofuels and coal to liquids.}\\

Unlike tight oil production in the USA, unconventional oil production from the tar sands in Canada has increased only slowly despite huge reserves of about 167 Gb.  It seems that production during 2013/2014 has reached a plateau around 2.2 mbd. In light of the past growth trend, and given the latest news about severe labor layoffs in the field, \cite{NYTsands}, we expect that Canada's tar-sand oil production might at most reach a plateau of 2.5 mbd by 2020 and remain at that level during coming decades.

Oil production from the Venezuelan oil sands is very hard to predict. In absence of clear data we assume that this production might begin in the next few years and eventually reach 1.2 mbd or about 50\% of the Canadian oil sand production. Keeping the very uncertain political future of Venezuela in mind, we assume that the country's oil-sand production might do no more than stabilise their overall oil production at around 2.5 mbd (section 3.5) from the near future to 2050. 

The production of bio fuels, \cite{EIAoildata}, is dominated by the USA (1.3 mbd) and Brazil (0.55 mbd), and even when taken together with liquids from coal contributed only about 2.7 mbd worldwide in 2014. As it seems rather unlikely that the global contribution from such fuels will change dramatically over coming decades, we assume that the production of such liquids will remain essentially unchanged at or below 3 mbd.

\subsection{Predicting 2014 production using the 2000 to 2005 data}

Assuming that the model described in this paper, had been suggested already in 2006, 
the corresponding predictions for the year 2014 can be compared with the available conventional 
crude oil production data from the EIA for the year 2014. 
This is especially interesting for Russia, China and the Middle East OPEC countries where 
economic (IEA) and resource based forecasts failed to describe the actual production trend (Table 1). 

Following the approach described at the beginning of section 3, we can try to use 
the corresponding three year average oil production for the years 2001* and 2004* to infer the 2006 production phase and make a forecast for the year 2014 (Table 4). As can be seen the model can be applied directly only 
for Western Europe, North-America and perhaps Asia+Oceania. For the other regions the beginning of the plateau 
and its length are difficult to infer because the production was growing in the other regions (see section 2).
However, as discussed in section 3, some political and economical factors, not free from wishful biases, can be used to 
estimate the end of the growth period which marks the year when our model can be used to make prediction for the year 2014. 

For the Persian Golf OPEC countries one would expect that the recovery from the 2003 Iraq war would lead to continued growth of about 1\%/year during the subsequent 5-10 years and a resulting production plateau around 23-25 mbd for several subsequent decades.

For Russia one could argue that the 3\%/year growth observed during those five years would lead production to a level close to the one that existed before the break up of the Soviet Union. Accordingly and following our model, we would have most likely predicted further growth leading to a 5 year production plateau at about 10 mbd until 2015, followed by the 3\%/year and 6\%/year decline periods.

{\tiny
\begin{table}[h]
\begin{center}
\begin{tabular}{|c|c|c|c|c|}
\hline
Region & production & production  & production     &modelled/production     \\
            & [mbd]         & [mbd/year]  & phase 2006  & production  [mbd]  \\
            & 2001*         &      2004*    &  [/year]          & 2014    \\ 
             \hline
\hline
Persian Golf (OPEC) &  19.5           &  21.3         & $\approx$ +1\%      &   23  / 24.3    \\
Russia                      &   6.9           &  8.6            & $\approx$ +3\%      &   10 / 10.1 \\
China                       &  3.3            &  3.5            & $\approx$ +3\%      &     4  / 4.2       \\
\hline
Western Europe         &    6.4          &   6.1           & $\approx$  -6\%     &     3.3 / 2.9        \\
Russia+Azer+Kaz.  &     8.1         &   10.2         & $\approx$  +5\%      &   12 / 12.6        \\
Asia+Oceania         &   7.45        &   7.5            & $\approx$  const     &     6.7 / 7.7       \\
Africa                      &    7.5          &   8.75         & $\approx$   2.5\%    &  10.3 / 8.0*     \\
South America        &    6.4          &   6.1           & $\approx$  -3\%      &   6.3  / 7.1        \\
North America        &    10.9        &   10.4         & $\approx$   -1\%      &   7.4  / 7.9       \\

\hline 
World (convent. oil)   &  68           &  71            &  $\approx$ 1\%        &  69  / 71 \\  
\hline 
\end{tabular} %\vspace{0.1cm}
\caption{Average conventional oil production 2001* (2000-2002) and  2004* (2003-2005), the inferred production phase and the modelled forecast and production for 2014.
}
\end{center}
\end{table}
}

For China, production was growing during the years prior to 2006 and, with the official reserves claimed in 2006, it looked reasonable to assume that production could increase by another few percent to reach a plateau of around 4 mbd between 2010 and 2015.

The actual production during the last 10 years in Russia, China and the Persian Golf OPEC countries is in reasonable agreement with the production predicted by our model.

In comparison, the 2006 World Energy Outlook (WEO 2006) overestimated the production increase from the Persian Golf region from 21 mbd (2004) to 26 mbd (2015) or about 2\%/year in comparison to the observed roughly 1\%/year growth rate. The IEA study correctly predicted the 2014 conventional oil production for Russia, but underestimated the possible production growth in China.

Some significant differences between our model prediction and the 2014 conventional oil production are 
observed for Africa and for the Americas. As has been discussed in the previous sections, 
the trends in Africa and South America were influenced by political and economic troubles,
which led to a lower production than predicted by our ``model", especially in Libya and other African countries. 
In contrast, and also influenced by the high oil prices, the exploitation of known oil fields in North and South America was seen during this period and resulted in a delay of the decline trend. But not even the high oil prices could stop the steep decline of oil production in Mexico. 
 
One might conclude that our model's predictions of 2010-2015 production, using production data from 2000 to 2005, were not only in reasonably good agreement with the actual regional and global conventional oil production trends observed during this period, but resulted in somewhat better predictions than the economic- and resources-based models did. But obviously, 
the real test for the predictive power of different forecasting models will be the comparison with the real production during the next 5-10 years. 
  
\subsection{A comparison with other 2015 models}
  
We conclude this section with a comparison of our model's forecast for the years 2020, 2030 and 2050 with the resource based forecast from J. Laherrr\`ere in April 2015 \cite{Laherrere} and with the November 2015 World Energy Outlook \cite{IEA2015} from the IEA (which includes predictions only through 2040).
  
Starting from a conventional oil production of about 71 mbd in 2014 and combining our 
modelled results from the different regions, we predict that the upper production limit will decline to 
66 mbd in 2020, 50 mbd in 2030 and 33 mbd in 2050. Adding all unconventional oil and oil-equivalent liquids, and including the 2014 refinery gains of about 2.5 mbd, we predict that an upper production limit for all liquids of about 93.5 mbd (2015) declining to 92.5 (2020),  followed by a decline to 79.5 mbd in 2030 and less than 62 mbd in 2050.

The forecasting method from J. Laherr\`ere is based on the Hubbert production profiles and the best available maximal crude oil reserve data. Laherr\`ere predicts a global conventional crude oil peak at about 73 mbd around 2015-2018. His estimates for 2020, 2030 and 2050 are about 72 mbd, 65 mbd and 35 mbd respectively.

When including all unconventional oil and oil-equivalent liquids, Laherr\`ere predicts a global production peak for all liquids (including refinery gains) around the year 2020 at approximately 94 mbd. This is followed by a decline to about 88 mbd in 2030 and 60 mbd around 2050.

The forecast from our model agrees with the calculation from Laherr\`ere with respect to the predicted possible production maximum and regarding the production around the year 2050. However, for the years 2020 and 2030 Laherr\`ereÕs production estimates are between 10 and 20\% higher than ours. The difference seems to originate primarily from  
Laherr\`ere's forecast of an increased production in the Persian Golf OPEC countries during the next decades compared to our forecast of a rather flat production.

Those two approaches can also be compared with the economic-based forecast from the IEA's World Energy Outlook 2015, \cite{IEA2015}.  Unfortunately, the definitions from the IEA and the EIA data (used as input for our model) for the various types of oil and oil-equivalent liquids are slightly different and the production numbers can differ by a few \%. To provide a better basis for the 
comparison between the different models, the IEA production numbers for the different years are multiplied 
by the ratio of the 2014 EIA and IEA production data. 
It is interesting to note that in contrast to the resource-constraint production models, the IEA2015 forecasts a rather constant conventional oil global production for the next several decades. Oddly, this forecast of a relatively constant level of production assumes a declining production of about 3.9\%/year,
significantly smaller than their own estimated 6\% average decline rate (WEO 2013), from existing fields combined with further exploitation of ``known" fields, exploitation of ``yet to be developed" fields and - especially - exploitation of ``yet to be discovered fields".

Taking only the forecasts for the existing fields and the further exploitation of the ``known" fields, the IEA prediction for conventional oil would thus be about 70 mbd in 2020, 62 mbd in 2030 and 52 mbd in 2040 and thus much closer to the resource-based forecasts.

\begin{figure}[h]
\begin{center}
\includegraphics[width=14cm]{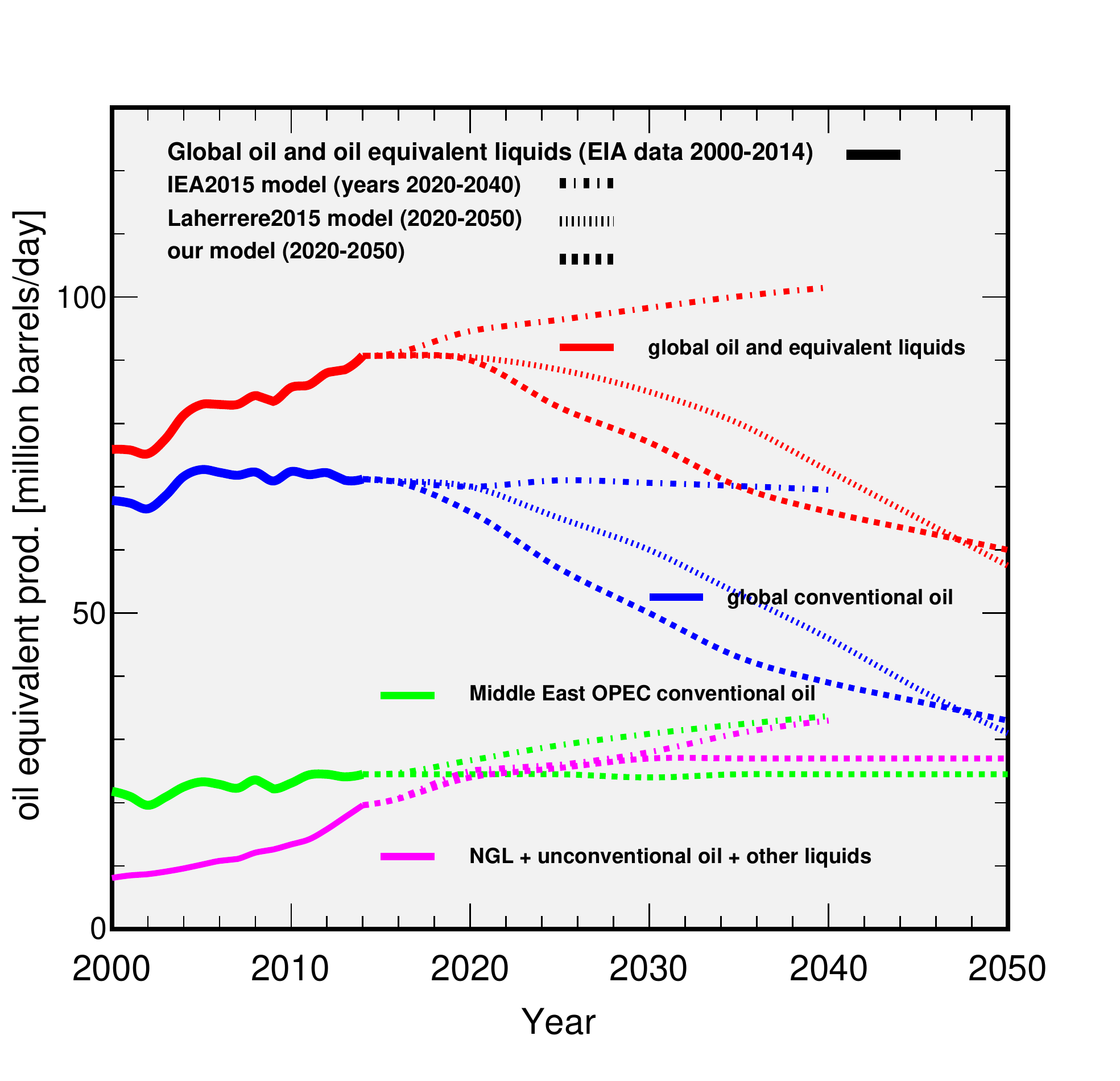}
\caption{\small{Global conventional crude oil, unconventional oil and oil-equivalent production data from 2000 to 2014 and projections from 2020 through 2050 as identified in the key at the top of the chart.}
}
\end{center}
\end{figure}

The striking difference between the IEA's predictions and the resource-based predictions is obvious. And it seems that those who take the resource-based models seriously should not only prepare for oil price volatility but should also prepare for changes in both the ratio of conventional oil to other petro-liquids and for increasing importance of the oil coming from the Persian Golf OPEC countries and how this oil might be distributed around the planet. Some ideas about near-term regional oil supply constraints will be discussed in the next section and more details about the regional oil supply situation will be presented in a subsequent paper (part II of this analysis).

\section{Near-term consequences for the large oil importers}

{\bf Oil exporting regions important to Western Europe}\\

Crude oil extraction in Western Europe declined from 6.2 mbd in the year 2002 to less than 2.87  mbd during 2014. It is interesting to note that this decline was only partially compensated with increasing imports, as the oil consumption in Western Europe was about 13 mbd in 2014, 2.5 mbd less than in the year 2004.  

With the very limited remaining reserves, it is rarely disputed that European oil production will continue its decline during the next decades. Following our model, we expect that by 2020 only about 1.9 mbd will be produced in Western Europe. 
During 2014 Western Europe had a net import of about 6 mbd, dominantly  through pipelines, from the FSU countries, Russia, Kazakhstan and Azerbaijan, about 2.1 mbd came from the Persian Golf OPEC countries and 2.1 mbd from Africa.
According to the IEA 2014 Medium-Term Oil Market report, \cite{IEAmidtermoil}, the IEA expects that the imports from these countries will drop by about 1 mbd (10\%) during the next 5 years. In contrast when following our modelled production declines,   
we expect that the total imports from the FSU countries will drop from 6 mbd today to about 4 mbd in 2020. 
Combining our modelled production decline estimates for Western Europe and the FSU countries, we expect that the oil consumption in Western Europe around the year 2020 will be about 3 mbd (between 20-25\% smaller than in 2014). \\

{\bf Oil exporting regions relevant for the USA}\\

During 2014 about 19 mbd of oil, about 10\% less than before the 2008 crisis year, 
were consumed in the USA.  According to the 2015 BP report, the net imports in 2014 were about 5.3 mbd of oil and oil-equivalent liquids. About 2.6 mbd and 0.3 mbd came from the direct neighbours Canada and Mexico respectively. The remaining net imports came from the Middle East (1.8 mbd), from South and Central America (0.3 mbd) and from Africa (0.3 mbd).  Assuming that the internal consumption in these regions remains roughly at today's numbers, we expect that the net imports from South America, Mexico and from Africa into the USA, about 1 mbd, will be terminated  during the next 5-10 years.
\\

{\bf Oil exporting regions relevant for the Asian Pacific countries} \\

Japan, South Korea and Singapore
are highly industrialised countries without domestic oil production. Their relatively high per capita oil consumption is thus 100\% imported. About 80\% of this oil, 6.5 mbd, comes from the Middle East.  
In contrast to Japan, where the oil consumption decreased during the last 10 years by 0.9 mbd, the consumption in South Korea an Singapore increased by 0.2 mbd and 0.5 mbd respectively. 
 
The other large oil importers are China and India. In 2014 China produced about 4.2 mbd of oil 
and had a net oil (and oil-equivalent liquids) import of about 7.1 mbd. The oil was mainly imported from the Persian Golf OPEC countries (3.5 mbd), from Russia and Kazakhstan (0.9 mbd), from West Africa (1.2 mbd) and from South America (0.75 mbd). 
India imported about 4 mbd, about 80\% of the consumer oil. This oil came dominantly from the Middle East (2.4 mbd), West Africa (0.6 mbd) and South America (0.7 mbd).

According to the BP 2015 report, in 2014 the remaining Asian Pacific countries imported another 4.8 mbd 
from the Middle East countries. Thus the Asian Pacific region as a whole imports more than 17 mbd from the oil rich Middle East countries. According to the IEA estimates, by 2020 the imports from this region will further increase by 0.8 mbd.  According to our estimate (section 3.7), and without war like disruptions, we expect a rather constant oil production in the OPEC Middle East countries and the next decades. As we expect that the oil production in Africa 
will start to decline during the next years we expect a steep decline of the oil exports 
from Africa.

\section{Summary}

In this paper we have presented a new approach to model the future oil production in different 
regions of the planet. This new approach is based on the regional conventional and unconventional oil
production trends observed during the past few years. 

Following the well documented oil production decline in Western Europe, Indonesia and Mexico 
during the past decade, we found that the observed oil production decline in these regions can be 
described with a very simple function. Once no significant new fields will be added to the production, the production stops growing and a plateau production might be maintained for several years. 
The length of the possible plateau production period depends on several factors, including the ability to export the additional oil from yet unexplored fields under profitable conditions.
However, for regions which have increased the production rapidly, we expect that 
the plateau production can be maintained only for about 5 years. Afterwards the production declines 
by about 3\%/year for a period of 5 years and followed by the terminal decline of 6\%/year. 

Applying this model for all oil producing regions of the planet, a forecast for the future regional oil
supply is obtained. We expect that only the Middle East OPEC countries might maintain their current production during the next decades. For all other regions we predict that 
the oil production will start to decline during the next few years. Knowing that especially the global ocean based transport system is constrained by the geography of the planet, it appears realistic that people in countries without significant domestic oil reserves, like many countries in Western Europe and Japan, will be confronted with the unavoidable declining oil availability and thus a less ``oil" based way of life during the next years.    \\
~~\\
\noindent

{\bf \large Acknowledgments\\} 

{\normalsize  \it{While the ideas presented are from the author alone, my thanks go to 
my friends, colleagues, and many students who have helped me during extensive exchanges to 
formulate these points and arguments. Especially I would like to thank W. Tamblyn for the critical and careful reading of the manuscript and for many suggestions on how to improve the clarity of the presented ideas. 
}}

\newpage
%\small


\begin{thebibliography}{99}
\bibitem{RussiaFSUEIABP}{Details about the oil production and exports in Russia and the other FSU 
countries can be found at the EIA website \cite{EIAdatabase}
and in the BP Statistical World Energy review \cite{BPstat15}.}
\end{thebibliography}
\end{document}